# Signatures of electronic phase separation in the Hall effect of anisotropically strained La$_{0.67}$Ca$_{0.33}$MnO$_3$ films

Liuqi Yu[1], Lingfei Wang[2], Xiaohang Zhang[3], W.B. Wu[2], S. von Molnár[1], Z. Fisk[4] and P. Xiong[1]

[1]Department of Physics, Florida State University, Tallahassee, Florida 32306, USA
[2]Hefei National Laboratory for Physical Sciences at Microscale, University of Science and Technology of China, Hefei 230026, China
[3]Department of Materials Science and Engineering, University of Maryland, College Park, Maryland 20742, USA
[4]Department of Physics, University of California, Irvine, California 92697, USA

**Abstract**

Systematic transport measurements have been performed on a series of La$_{0.67}$Ca$_{0.33}$MnO$_3$ (LCMO) thin films with varying degrees of anisotropic strain. The strain is induced via epitaxial growth on NdGaO$_3$(001) substrates and varied by controlling the thermal annealing time. An antiferromagnetic insulating (AFI) state, possibly associated with charge ordering, emerges upon thermal annealing. The Hall effect in these materials exhibits features that are indicative of a percolative phase transition and correlate closely with the emergence of the AFI state. In the paramagnetic phase, the Hall resistivity takes on two slopes in all samples: a decreasing negative slope with increasing temperature at low fields, which is attributed to the carrier hopping motion, and an almost temperature independent positive slope at high fields due to diffusive transport of holes. Significantly, the crossover fields of the Hall resistivity slope at different temperatures correspond to the same magnetization, which is interpreted as the critical point of a magnetic field-driven percolative phase transition. At lower temperatures near the zero-field metal-insulator transition, pronounced enhancement of the Hall coefficient with the development of the AFI state is observed. The enhancement peaks near the magnetic field-driven percolation; its magnitude correlates with the strength of the AFI state and is suppressed with the melting of the AFI state by an in-plane magnetic field. The observations resemble many features of the enhancement of the Hall coefficient in granular metal films near the composition-driven percolation.

## 1. Introduction

The observation of colossal magnetoresistance (CMR) in doped manganites near the metal-insulator transition [1, 2] has led to a dramatic revival of interest in this class of perovskite materials. Since then, extensive efforts in both theory and experiment have been expended to understand the correlation of the rich electronic transport behavior with complex magnetic, lattice, charge, and orbital ordering and their intricate interplay [3, 4]. The doped manganites have a generic composition of $Re_{1-x}A_xMnO_3$, where Re is a trivalent rare earth ion and A a divalent alkaline earth ion, leading to a mixed valence for Mn. In many of the 'optimally' doped manganites, the double exchange (DE) [5-7], namely, the interaction of the outer-shell orbitals of $Mn^{3+}$ and $Mn^{4+}$ through the O(2p) orbital, which promotes ferromagnetism, results in a metallic ground state at low temperatures. Near and above the Curie temperature ($T_C$), a strong electron-phonon coupling due to the Jahn-Teller (JT) [8, 9] effect leads to localization of the $e_g$ itinerant electrons on the $Mn^{3+}$ site. It is now widely accepted that the DE interaction and JT distortion are necessary ingredients to the ferromagnet-paramagnet and metal-insulator transitions in these materials.

On the other hand, at doping levels close to rational numbers and/or with appropriate combinations of the rare earth and alkaline ions (e.g., Pr and Ca), the doped manganites can develop long-range ordering of the cations of different valences below the Curie temperature, i.e., charge ordering (CO) [10, 11]. CO also competes with DE, leading to an antiferromagnetic insulating (AFI) ground state. For example, in $Pr_{0.65}Ca_{0.35}MnO_3$, a steeply rising resistivity with decreasing temperature appears below a characteristic ordering temperature, $T_{CO}$ [12]. A sufficiently large magnetic field can lead to the melting of the CO and a ferromagnetic metallic (FMM) state, which manifests in CMR below $T_{CO}$ [12]. The optimally doped $La_{2/3}Ca_{1/3}MnO_3$ (LCMO) is a prototypical material belonging to the first type (with a FMM ground state). However, it has been demonstrated that an AFI state [13, 14], possibly due to short-range CO [13], can be induced in ultrathin LCMO films by introducing an in-plane anisotropic strain field from the substrate. The approach of using substrate induced strain to modulate the magnetic and electronic states of ultrathin epitaxially grown films has been widely applied to a variety of manganite materials, especially those doped into or on the verge of having significant AFI ordering [15, 16].

Although these essential pieces of physics (DE, JT, CO) had been identified for some time, it was demonstrated both theoretically and experimentally that a full account of the phase transitions and the associated CMR cannot be achieved without invoking electronic phase separation (EPS) [17]. There is extensive experimental evidence for the existence of mesoscale inhomogeneities (from nm to μm) in both types of materials with either FMM [18, 19] or CO [20] ground state. The insulator-metal transitions, driven by temperature or magnetic field, are consequently percolative in nature. Historically, similar mesoscale heterogeneities and magnetically driven percolative transitions had been observed in rare earth chalcogenides [21, 22] and hexaborides [23]. In the chalcogenides, the conducting ferromagnetic clusters in the insulating paramagnetic/antiferromagnetic background were identified as bound magnetic polarons [22, 24].

Experimentally, EPS has been probed by a wide variety of direct (e.g., scanning tunneling microscopy [18, 19, 25] and electron microscopy [20]) and indirect (e.g., noise spectroscopy [26, 27], single domain resistance fluctuation [28, 29], and neutron scattering [30-32]) techniques in the manganites. In the study of the magnetically driven EPS in the semimetallic ferromagnet $EuB_6$, we recently discovered an intriguing manifestation of the EPS and percolative phase transition in the nonlinear Hall effect (HE) [33]: We observed a distinct switch in the slope of the Hall resistivity of $EuB_6$ in its paramagnetic phase, which evolves systematically with temperature. The switching field depends linearly on temperature and extrapolates precisely to the Curie-Weiss temperature of the material, which indicates that the switches in the HE slope occur at a *single constant critical magnetization* over a wide temperature range. We interpret the critical magnetization as the point of percolation for patches of a more conducting and magnetically ordered phase in a less ordered background. Moreover, this picture appears applicable in the paramagnetic phase of $Nd_{0.7}Sr_{0.3}MnO_3$ [33, 34], although overall the HE in the mixed-valence manganites is far more complex [35-38]. The HE therefore could provide a readily available tool for probing EPS.

Here we report systematic measurements of the magnetic field and temperature dependencies of the Hall resistivity, especially close to $T_C$, in three samples of LCMO. All three samples were at the optimal Ca doping of 1/3 and grown under identical conditions on the same $NdGaO_3$(001) substrates, but subject to different thermal annealing (varying lengths of annealing time at the same temperature). Thus the phase evolution in the samples is driven by a single parameter, effectively eliminating the complications associated with varying the doping level. The Hall effect in these materials exhibits two prominent features reflective of a percolative phase transition in an increasingly insulating background. 1) In the paramagnetic phase, the Hall resistivity exhibits a change of slope whose crossover fields at different temperatures correspond to the same magnetization, similar to the observation in $EuB_6$ [33]. 2) In the vincinity of $T_C$, a pronounced enhancement of the Hall coefficient near the percolation field develops with increasing annealing time. The magnitude of the enhancement grows with increasing strength of the AFI state and is suppressed by an in-plane magnetic field induced melting of the AFI state. The enhancement of the Hall coefficient here is reminiscent of the giant Hall effect in granular films of a percolating metallic network in an insulating matrix near the composition-driven percolation threshold [39], thus is an additional manifestation of a magnetically driven percolative transition.

## 2. Experiments

Epitaxial LCMO ($a_C$ = 3.864 Å) thin films of 45 nm thickness were grown on NGO(001) ($a_C$ = 3.867 Å) by pulsed laser deposition at fixed temperature and oxygen pressure of 780 $^o$C and 45 Pa respectively, and subsequently annealed *ex situ* at 780 $^o$C for 1 hour, 10 hours and 20 hours. Flowing $O_2$ gas was used during the annealing to avoid any structural changes due to oxygen loss. All three samples were patterned into a Hall-bar geometry by standard photolithography and wet-etching in a $HPO_3$ (3.4%) / $H_2O_2$ (1.5%) acid solution. The magnetoresistance and Hall effect measurements were carried out in a $^4$He cryostat with the sample temperature ranging from 5 K to 300 K. A magnetic field up to 8 T and a DC current from 0.25 mA to 0.5 mA were used for the HE measurements. I-V measurement was carried out before each Hall measurement in order to ensure that the applied current in the Hall measurement is in the linear region and the effect of Joule heating is negligible.

## 3. Results and Discussion

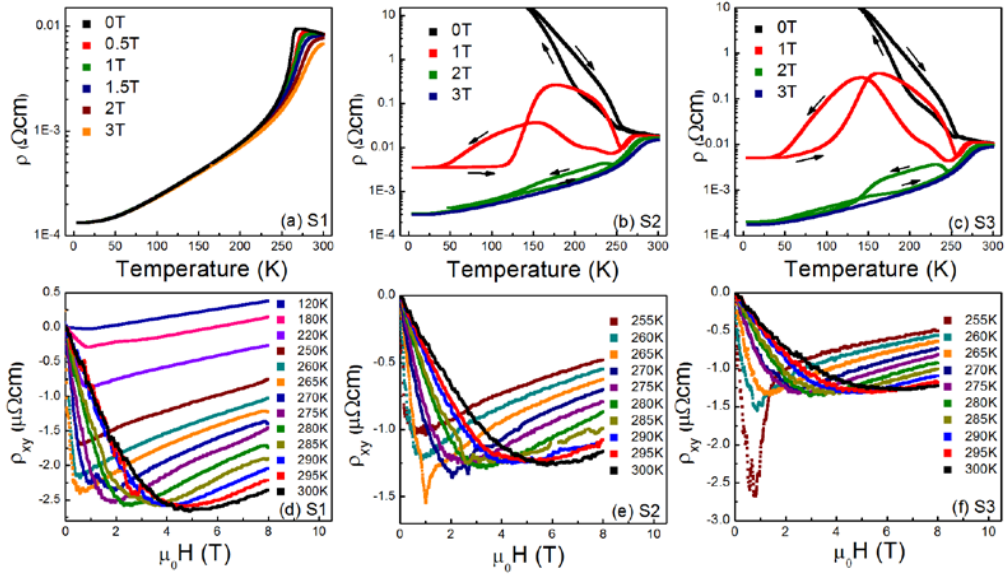

Figure 1. (a)(b)(c) Resistivity as a function of temperature for three identically grown samples of 45 nm thick LCMO on NGO and post-annealed at 780°C in flowing $O_2$ for 1 hour, 10 hours and 20 hours. (d)(e)(f) Hall resistivity of the three samples as functions of applied magnetic field at selected temperatures.

Figure 1 shows the temperature dependence of the longitudinal resistivity measured at selected magnetic fields (top panels) and the magnetic field dependence of the Hall resistivity at selected temperatures (bottom panels) for the three LCMO/NGO samples: sample 1 (S1, annealed for 1 hour), sample 2 (S2, annealed for 10 hours) and sample 3 (S3, annealed for 20 hours). The overall behavior and evolution of $\rho(T)$ of the three samples are consistent with results from previous studies with tuning parameters of thickness and annealing temperature or time [13, 14]. It is shown recently that the annealing promotes better coherency with the substrate and enhances the anisotropic strain, thus creating and strengthening the AFI state [40]. Here it is evident that an AFI state is created by the thermal annealing in S2 and S3, as indicated by the sharply rising resistivity with thermal hysteresis below a characteristic temperature, $T_{CO}$. Microscopically, the emergence of the AFI state has been attributed to strain-induced charge/orbital ordering. The application of a magnetic field could suppress such ordering and restore the FMM state.

The temperature and field dependencies of the resistivity [Figure 1(a)] and Hall resistivity [Figure 1(d)] of S1 are typical of bulk LCMO reported extensively before. It shows the paramagnet-ferromagnet and metal-insulator transitions, as evidenced by the zero-field resistivity peak at around 270 K, which is suppressed by applying a magnetic field, showing the CMR effect around the transition region. Figure 1(d) shows the overall features of the Hall resistivity for S1 at various temperatures across $T_C$. As in bulk or unstrained films of LCMO, in the ferromagnetic phase, the Hall resistivity exhibits a negative component at low field and a positive high-field Hall slope. The high field component is believed to arise from the ordinary Hall effect due to diffusive motion of the holes in the metallic state [35, 36]. With increasing temperature, the slope gradually increases, suggesting a decrease in the effective carrier density. This may be an indicator of the growing volume fraction of the insulting state when approaching $T_C$. Also with increasing temperature, the low-field negative component emerges and becomes increasingly significant. This component was found to scale with the magnetization and determined to be the anomalous HE well into the ferromagnetic state [35, 38]. In the paramagnetic phase, the Hall resistivity also takes on a negative low-field slope which decreases with increasing temperature, before reverting to a positive slope at high fields similar to the one at low temperatures. However, the low-field Hall component here is related to neither the conventional HE of electrons nor the anomalous HE [38]; rather, it is attributed to the hopping motion of the charge carriers: Well into the paramagnetic state (typically $T > 1.4T_C$), the low-field Hall slope follows an activation behavior of the small polaron hopping [41-43], which is consistent with the hopping mechanism first proposed by Friedman and Holstein [44, 45]. Closer to $T_C$, the microscopic picture of the HE becomes more complex and less understood. The expected activation behavior does not extend to the vicinity of $T_C$, and the hopping conduction and the low-field HE are strongly influenced by the proliferation of EPS and the inhomogeneous magnetic state [42, 43]. Our samples have $T_C$ from 258 K to 263 K, which confines our high temperature HE results to the latter regime.

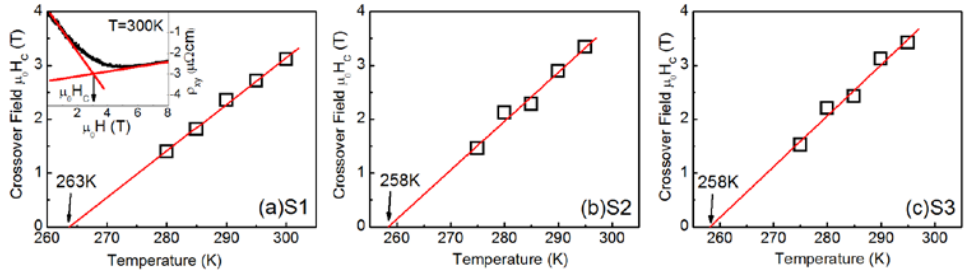

Figure 2. (a)(b)(c) Temperature dependence of the crossover field of the Hall resistivity for S1, S2 and S3 respectively. Inset: an example (T = 300 K) illustrating the determination of the crossover field.

We first analyze the Hall resistivity in the paramagnetic phase following our previous study of the semimetallic ferromagnet $EuB_6$ [33, 46]. Linear fits to the low-field and high-field Hall resistivity are performed for each temperature well above $T_C$, as exemplified in the inset of Figure 2(a). The intersection of the two best-fit lines is identified as the crossover field $H_C$ which depends linearly on temperature as shown in Figure 2. For all three samples, $H_C$ extrapolates to a temperature very close to the paramagnetic Curie temperature $\theta$, which is normally slightly lower than the corresponding metal-insulator transition temperature [47]. Since in the paramagnetic phase, $H/(T-$ $\theta$) is proportional to magnetization, the results in Figure 2 indicate that for each sample in the entire measurement temperature range, the transition from the polaronic hopping to diffusive transport occurs at a single critical magnetization. Evidently, this fitting holds for all three samples, suggesting that they share a similar physical origin for the nonlinear HE in the paramagnetic state despite the different degrees of anisotropic strain, and the emergent AFI state only affects the HE results at lower temperatures.

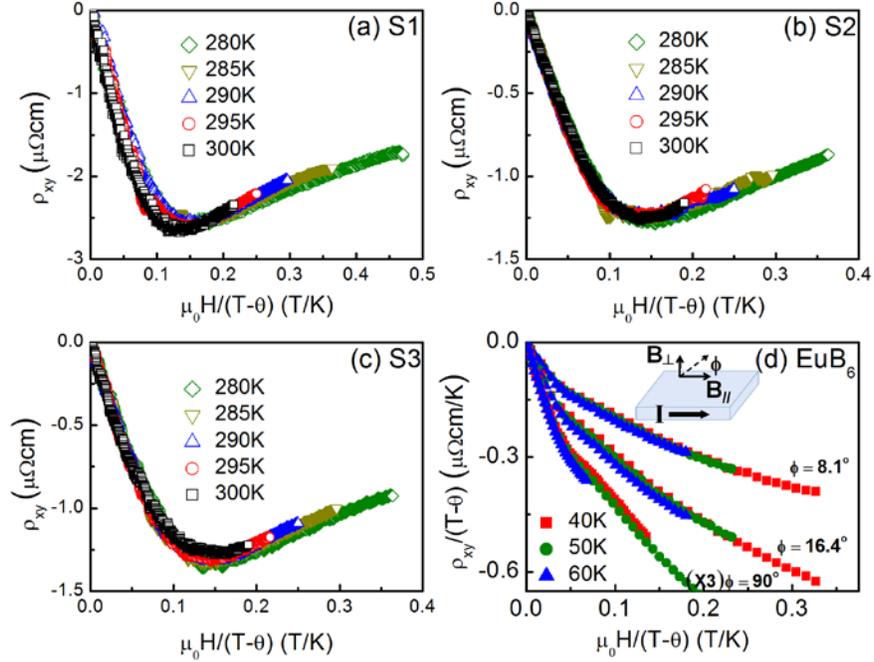

Figure 3. (a)(b)(c) Hall resistivities of the LCMO samples well into the paramagnetic state, where the independent variable, $H/(T-\theta)$, is proportional to the magnetization. (d) Scaled Hall resistivity versus effective magnetization for a $EuB_6$ crystal. The total field is used to calculate the effective magnetization. In both cases, the Hall resistivity changes slope at a single critical magnetization regardless of temperature and field orientation.

Figure 3 presents direct evidence of the Hall slope switching at a constant magnetization by plotting the Hall resistivity as a function of $H/(T-\theta)$, which is directly proportional to the magnetization in the paramagnetic state. The results for the LCMO samples, and for comparison, a sample of $EuB_6$, are plotted. In the case of $EuB_6$ [Figure 3(d)], the Hall resistivities are rescaled by the factor $1/(T-\theta)$ ($\theta$ = 15.6 K) in order to retain the slope of $\rho_{xy}$ versus $H$. The Hall measurements here were performed deliberately with the magnetic field oriented at different angles with respect to the sample plane; it is evident that, regardless of the direction of the applied magnetic field, the Hall slope switchings occur at the same value of $H/(T-\theta)$ where $H$ is the *total* field. Figures 3(a)-(c) show the Hall resistivities of the LCMO samples, which collapse onto a single curve, suggesting a singular dependence of the Hall resistivity on magnetization in this temperature range. The observation is consistent with the scaling behavior with magnetization observed in several manganite materials [42, 47]. It is important to note the common features revealed in Figures 3(a)-(c) and Figure 3(d), despite the significant differences in the microscopic pictures of the HE in LCMO and EuB$_6$, especially the interpretation of the low-field Hall slope: The low-field Hall slope of EuB$_6$ is predominantly from the diffusive electrons [33, 46], while in LCMO it is a result of polaronic hopping. In both magnetic systems, the percolative nature of the transport and phase transitions is clearly reflected in the HE.

We now turn to the HE near the transition temperature: In S2 and S3, the most notable difference in the Hall resistivity is the appearance of a peak in its (negative) magnitude [Figure 1(e)(f)]. It is apparent in S2 and becomes much more pronounced in S3. Comparing the resistivity and Hall data in Figure 1, it is evident that the appearance and evolution of the Hall resistivity dips correlate closely with the emergence and strengthening of the AFI state with increasing annealing time. To provide direct evidence for this correlation, we note a strong anisotropy in the melting of the AFI state: It is easier to suppress the AFI state with an in-plane magnetic field than a perpendicular one. As indicated by the $\rho(T)$ data in Figure 4(a), the AFI state in S3 is almost fully suppressed by an in-plane field of 2 T but still present under the same out-of-plane field. We therefore performed HE measurements on S3 with the magnetic field oriented at 44$^o$ relative to the sample plane. As shown in Figure 4(c), the Hall resistivity dips vanished in the tilted field and there is a direct transition from a negative low-field slope to a constant positive high-field slope. This is in striking contrast to the data in perpendicular field [Figure 4(b)], where the negative slope component reaches a much larger (absolute) value *with an abruption transition to a steep positive slope* which decreases with increasing field and eventually saturates at a constant value at high fields. These differences are most likely due to a more significant melting of the AFI state by the in-plane component of the magnetic field in the case of the tilted applied field.

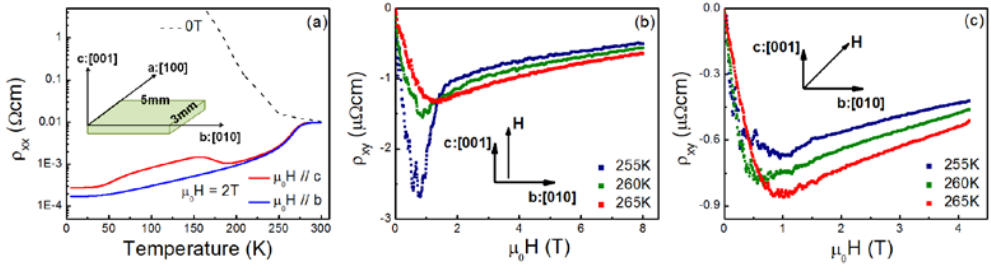

Figure 4. (a) Temperature dependence of the longitudinal resistivity of S3 in zero magnetic field (dotted lines) and 2 T (solid lines) magnetic field along the c-axis (red) and b-axis (blue). (b)(c) HE of S3 with the magnetic field perpendicular and tilted at 44$^o$ with respect to the sample plane. The field direction is determined with a commercial Hall sensor.

The experimental results in Figure 1 and Figure 4 point to an essential ingredient for the dip feature in the Hall resistivity: a robust zero-field AFI state which provides a sufficiently insulating electronic background. In S2 and S3, the highly insulating background makes the system more closely resemble a film of percolating granular metal in an insulating matrix. In systems of metal-insulator composite with a metallic percolating network, classical percolation theory expects an enhancement of the (ordinary) Hall coefficient near the *classical* (geometrical) percolation threshold ($x_c$) based on the reduction of the effective carrier density from decreasing metal volume fraction [48]. Since the mid-1990s, giant Hall effect (GHE) has been reported in a variety of magnetic [49-52] and nonmagnetic [39, 53] granular metal films; an enhancement of the Hall coefficient as large as $10^3$ is observed. The enhancement is much larger than the prediction of the classical percolation ($< 10^2$), and it peaks at the *quantum*, rather than classical, percolation threshold, as determined by the more detailed recent experiments [39, 53]. In the quantum percolation model, in a narrow region above the geometrical percolation threshold, multiple scattering of the electronic waves in the random percolating conducting channels leads to localization of the electrons and thus overall nonmetallic behavior [54, 55]. Global metallic transport is not achieved until a higher metal volume fraction, $x_q$, namely the quantum percolation threshold. Wan and Sheng [56] proposed that in the region $x \sim x_q$ and when the electron dephasing length ($L_\Phi$) is larger than the granular structure feature size (s), local quantum inference over the length scale of $L_\Phi$ leads to a significant reduction of the effective carrier density, and equivalently, a giant increase of the Hall coefficient, of a macroscopic sample.

The EPS and resulting percolative phase transition have profound impacts on the HE of all three samples near $T_C$. In all three samples, the magnetic field drives the proliferation and eventual percolation of the FMM regions, however, the manifestation of this process in the HE is quite different in different samples. To illustrate this, we chose one Hall trace for each sample at a representative temperature which exhibits the most pronounced nonlinearity (Hall trace at 265 K for S1, 265 K for S2 and 255 K for S3), and obtained numerically the slope as a function of the applied field. At each Hall resistivity minimum, there is a rather sudden change of slope from a negative value to a large positive value, in contrast to the cases deeper into the paramagnetic state (higher temperatures) where the slope changes continuously from negative to positive (e.g., inset of Figure 2(a)). In Figure 5, we take the field at which the slope changes sign as the point of percolation and plot the positive slope (i.e., the Hall coefficient) beyond percolation on a logarithmic scale. Right beyond percolation the Hall slope of S3 attains a very large positive value, which decreases sharply with increasing field before a high field region of gradual decrease. Similar, albeit weaker, features are seen in S2. Even in S1, there exists a pronounced nonlinear region right above percolation, which manifests in the increasing Hall slope with decreasing field toward percolation. Similar nonlinearity as in S1 has been widely observed in unstrained LCMO films and crystals in the transition region (see, for example, Fig. 1 of Ref. 38), and is expected from a percolative phase transition: Beyond the percolation point, increasing magnetic field leads to larger volume fraction for the FMM phase, thus larger effective hole density (smaller Hall coefficient) for the sample. In S2 and S3, the high-field behavior is similar to that in S1, however, close to percolation ($< 2$ T) there is a significant enhancement of the Hall coefficient.

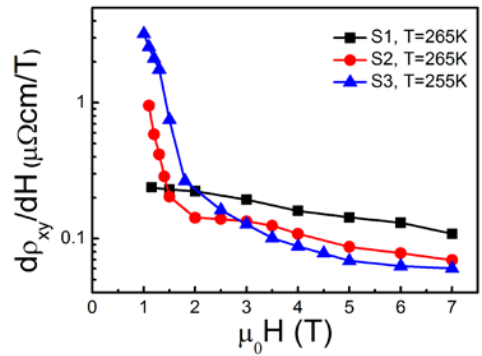

Figure 5. Derivative of a selected Hall resistivity trace for each sample beyond percolation point. Each chosen Hall resistivity trace has the most pronounced nonlinearity for the corresponding sample.

The magnitude of the HE enhancement near percolation appears to depend on how insulating the electronic background is and how sensitive it responds to the external magnetic field. It is also worth noting that in S3, with a tilted applied field (Figure 4), the HE enhancement is greatly diminished at 255 K and completely eliminated at other temperatures where we see a direct transition from negative Hall slope at low fields to a constant, positive slope at high fields. These observations suggest a close correlation between the enhancement of the Hall coefficient near percolation in S2 and S3 and the presence and strength of the AFI state in the samples. They also point to a probable connection with the enhancement of the Hall coefficient in granular metal films. Presently, the experimental evidence is not sufficient to discern whether the observed Hall coefficient enhancement results from classical or quantum percolation, namely, whether the mechanism of local quantum interference plays a significant role. Considering the rather high temperature and short electron mean free path (at the order of lattice spacing), the electron dephasing length is likely to be short in these materials. On the other hand, the feature sizes of the electronic inhomogeneities can also be very small (~12 Å [30]). Regardless of the exact origin, the result reveals an important manifestation of the magnetic field-driven percolative phase transition in the Hall effect of the manganites with an insulating ground state. It would be interesting to experimentally determine the key relevant length scales in the systems in order to decipher the role, if any, of the quantum interference, which could provide a unique probe for the EPS and its evolution.

## 4. Conclusion

To summarize, by controlling the anisotropic strain via thermal annealing in a set of identically grown LCMO samples, AFI state of increasing strength is induced without the complications from varying compositions. Two prominent nonlinear features in the HE of these samples are identified as manifestations of EPS and magnetically driven percolation: In the paramagnetic phase, regardless of the different ground states at low temperature, all three samples exhibit similar switch from a negative Hall slope to a positive one, which is found to occur at the same critical magnetization in the temperature range. Approaching the Curie temperature, large enhancement of the Hall coefficient near the percolation field is observed in strained samples, which correlates with the emergence of the AFI state and a more insulating electronic background. The pronounced enhancement of the Hall coefficient is reminiscent of the GHE observed in granular metal films near the composition-driven percolation thresholds. The resemblance to the physically inhomogeneous and percolating systems provides strong evidence that it originates from the EPS and magnetic field-driven percolation. The prevalence of the first nonlinear HE feature in a variety of strongly correlated materials may be an effective and broadly applicable tool for probing EPS induced by magnetic correlation.

**Acknowledgment**

We acknowledge valuable discussions with Abdelmadjid Anane and Jens Müller. This work was supported by NSF under grants DMR-0908625 and DMR-1308613.